\documentclass[letterpaper, 10 pt, conference]{ieeeconf}  

\usepackage{graphicx}
\usepackage{amsmath,amssymb,amsfonts}
\usepackage{multirow}
\usepackage{algorithm}
\usepackage{algorithmic}
\usepackage{psfrag}
\usepackage{subfigure}
\usepackage{color}
\usepackage{soul}
\usepackage{xcolor}
\usepackage[bookmarks=false]{hyperref}
\usepackage{lipsum}
\usepackage{mathrsfs}
\usepackage{pbox}
\usepackage{flushend}
\usepackage{mathtools}

\makeatletter
\def\hlinewd#1{%
  \noalign{\ifnum0=`}\fi\hrule \@height #1 \futurelet
   \reserved@a\@xhline}
\makeatother

\graphicspath{{Figures/}}

\IEEEoverridecommandlockouts                              

\title{\LARGE \bf
LCS-TF: Multi-Agent Deep Reinforcement Learning-Based Intelligent Lane-Change System for Improving Traffic Flow
}

\author{Lokesh Chandra Das$^{1}$ and Myounggyu Won$^{1}$
\thanks{$^{1}$Lokesh Chandra Das and Myounggyu Won are with the Department of Computer Science, University of Memphis, Memphis, TN, United States
        {\tt\small \{ldas, mwon\}@memphis.edu}}%
}

\begin{document}

\maketitle
\thispagestyle{empty}
\pagestyle{empty}

\begin{abstract}
Discretionary lane-change is one of the critical challenges for autonomous vehicle (AV) design as it has a significant impact on traffic efficiency, safety, and driving comfort. Most of the existing intelligent lane-change solutions have primarily focused on optimizing the performance of the ego-vehicle, thereby suffering from limited generalization performance. Recent research has seen an increased interest in multi-agent reinforcement learning (MARL)-based approaches to address the limitation of the ego vehicle-based solutions through close coordination of multiple agents. Although MARL-based approaches have shown promising results, the potential impact of lane-change decisions on the overall traffic flow of a road segment has not been fully considered. Moreover, in many existing solutions, the driving safety and comfort models incorporated in the reward function are oversimplified, potentially leading to suboptimal performance.

In this paper, we present LCS-TF, a novel hybrid MARL-based intelligent lane-change system for AVs designed to jointly optimize the local performance for the ego vehicle, along with the global performance focused on the overall traffic flow of a given road segment. With a careful review of the relevant transportation literature, a novel state space is designed to integrate both the critical local traffic information pertaining to the surrounding vehicles of the ego vehicle, as well as the global traffic information obtained from a road-side unit (RSU) responsible for managing a road segment. We create a reward function to ensure that the agents make effective lane-change decisions by taking into account the performance of the ego vehicle and the overall improvement of traffic flow. A multi-agent deep Q-network (DQN) algorithm is designed to determine the optimal policy for each agent to effectively cooperate in performing lane-change maneuvers. LCS-TF's performance was evaluated through extensive simulations in comparison with state-of-the-art MARL models. In all aspects of traffic efficiency, driving safety, and driver comfort, the results indicate that LCS-TF exhibits superior performance.
\end{abstract}


\section{Introduction}
\label{sec:introduction}


Autonomous driving has garnered significant attention from both the research community and industry, owing to its high potential for reducing traffic congestion and enhancing driving safety~\cite{wang2021harmonious,zhou2022multi}. Discretionary lane change for autonomous vehicles (AVs) is a key research focus in autonomous driving due to its substantial impact on both traffic efficiency and driving safety~\cite{nilsson2016lane,zheng2019cooperative,ge2022heterogeneous}. Especially in dense or congested traffic environments, relying solely on the longitudinal motion control of AVs is insufficient for enhancing traffic flow. Timely and agile lane changes are critical to improve traffic efficiency~\cite{li2022decision}. However, making effective lane-change decisions for AVs can be particularly challenging, especially in complex and dynamic traffic environments where AVs and human-driven vehicles (HVs) coexist~\cite{chen2020autonomous}. 

A substantial body of research has been conducted on lane-changing systems for AVs. The existing research works on lane-changing systems for AVs can be broadly classified into rule-based and machine learning-based approaches. Rule-based approaches rely on hand-crafted rules designed to emulate the behaviors of human drivers~\cite{ahmed1996models,sun2012lane}. However, despite their usefulness for specific scenarios, rule-based approaches suffer from limited generalization performance~\cite{zhang2022learning}. In response to the limitations of rule-based approaches, machine learning-based approaches have been the subject of active research especially concentrating on reinforcement learning (RL)-based methods~\cite{du2020cooperative,chen2020autonomous,li2022combining}. 

We classify the RL-based solutions into two main categories: the ego vehicle-based approaches~\cite{wang2018reinforcement,ye2020automated,zhang2022learning,he2022robust} and multi agent-based approaches~\cite{ha2020leveraging,chen2021graph,zhou2022multi,hou2021decentralized,zhang2022multi}. Ego vehicle-based approaches aim to enhance the efficiency and safety of the ego vehicle by incorporating surrounding vehicles as part of the state of the RL model. While ego vehicle-based approaches are highly scalable, they may result in poor performance since the complex interaction between vehicles is not effectively accounted for. In contrast, the MARL-based solutions exploit the close coordination of multiple agents to reach better lane-changing decisions for multiple autonomous vehicles, leading to higher traffic efficiency and safety. However, the majority of existing MARL-based solutions rely on onboard sensor data to make lane-changing decisions, thereby restricting the perception capability of the agents to the surrounding area~\cite{zhang2022multi,wang2021harmonious}. More importantly, the potential of intelligent lane-changing systems for improving the overall traffic flow of a road segment has not been fully leveraged.

In this paper, we propose a novel MARL-based intelligent lane-change system for AVs, namely LCS-TF, that simultaneously optimizes the efficiency, safety, and driving comfort of the ego vehicle, along with the traffic flow of the road segment to which the agents belong. To the best of our knowledge, this is the first study to present a hybrid intelligent lane-change approach that maximizes both local and global performance through the incorporation of a road-side unit (RSU) that manages a road segment, and vehicle-to-everything (V2X) capabilities for the agents. Specifically, the proposed intelligent lane-change problem is formulated as a decentralized partially observable Markov decision process. We then develop a MARL framework to solve the problem. Our state space design incorporates both local traffic information obtained from surrounding vehicles of an ego vehicle, as well as global traffic information collected from the RSU responsible for managing the corresponding road segment. A novel reward function is created to effectively carry out the joint optimization for the performance of the ego vehicle and the overall improvement of traffic flow. A multi-agent DQN algorithm is designed to identify the optimal policy that each agent can use to collaboratively execute lane-change maneuvers. Extensive simulations were conducted to evaluate the performance of LCS-TF in comparison with two main types of existing MARL models~\cite{zhang2022multi,zhou2022multi}. The results demonstrate the superior performance of LCS-TF in all aspects of traffic efficiency, driving safety, and driver comfort.

This paper is organized as follows. We present a review of the latest efforts for intelligent lane-change solutions for AVs in Section~\ref{sec:related_work}. We then present the details of the proposed intelligent lane-change solution in Section~\ref{sec:proposed_approach}. In Section~\ref{sec:experiments}, we present the simulation results. We then conclude in Section~\ref{sec:conclusion}.

\section{Related Work}
\label{sec:related_work}

\begin{figure*}[h]
	\centering
	\includegraphics[width=.99\textwidth]{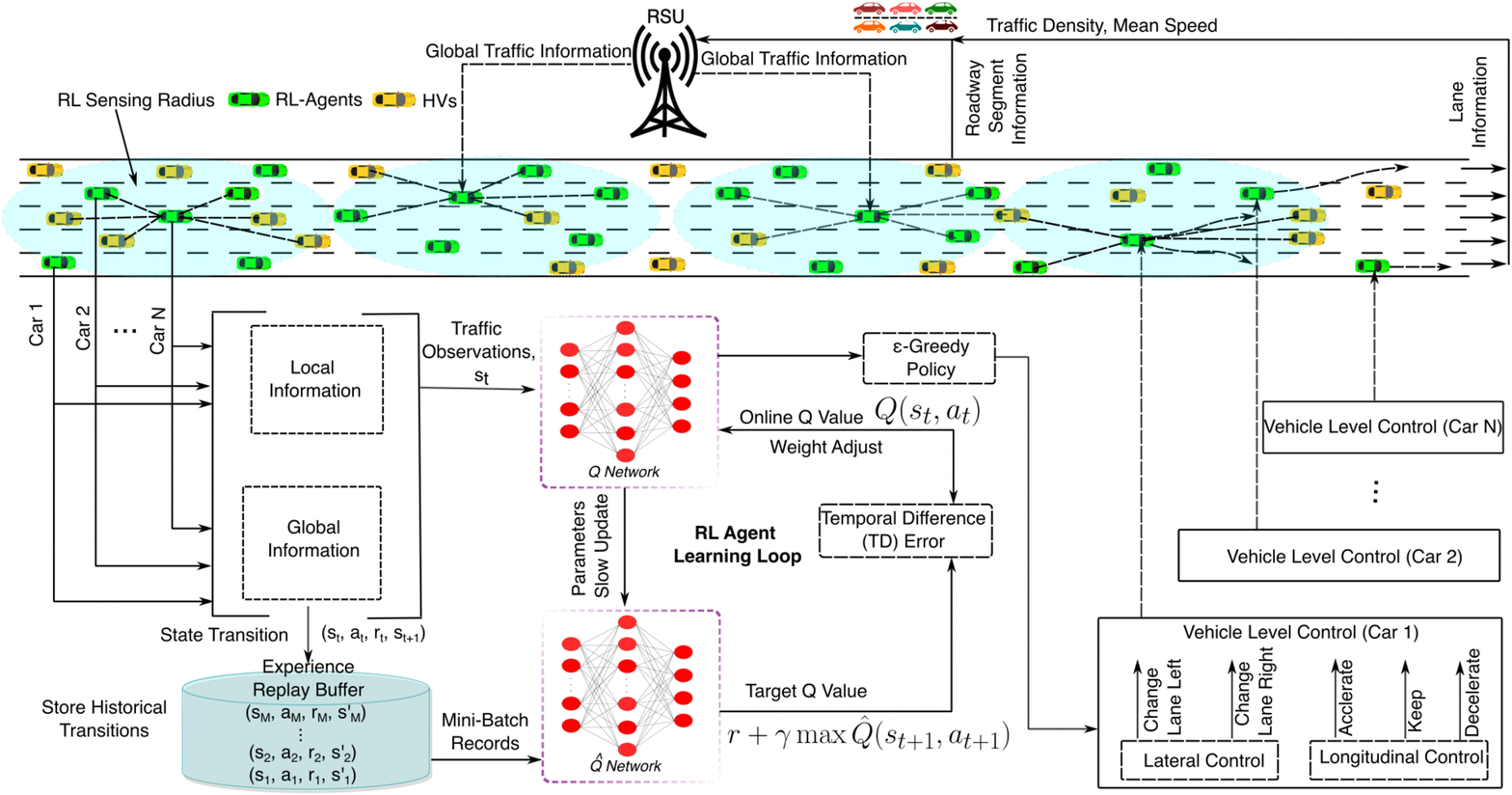}
	\caption {An overview of the operation of LCS-TF.}
	\label{fig:overview}
\end{figure*}

This section presents a review of machine learning-based intelligent lane-changing systems for autonomous vehicles. Most of the existing works are concentrated on improving the safety and efficiency of the ego vehicle~\cite{wang2018reinforcement,ye2020automated,zhang2022learning,he2022robust}. These ego vehicle-based approaches aim to train the lane-changing behavior of the ego vehicle in a singe-AV setting where the surrounding vehicles are considered as the environment for decision making~\cite{du2020cooperative,chen2020autonomous,li2022combining}. 

Wang \emph{et al.} proposed an RL framework to train the ego vehicle for intelligent lane-changing in diverse traffic environments~\cite{wang2018reinforcement}. In particular, a Q-function based on a closed form greedy policy and a deep Q-learning algorithm based on continuous state and action space were designed to develop the RL framework. Ye \emph{et al.}'s work was focused on improving the learning efficiency~\cite{ye2020automated}. They created a lane-changing strategy based on proximal policy optimization-based deep RL and demonstrated that stable performance could be achieved with improved learning efficiency. Zhang \emph{et al.} took into account human factors for training the ego vehicle~\cite{zhang2022learning}. The driving styles of both the ego vehicle and surrounding vehicles were incorporated into their decision making model. He \emph{et al.} concentrated on improving the robustness of their machine learning model for intelligent lane-changing systems~\cite{he2022robust}. They proposed an observation adversarial RL approach with an aim to manage observation uncertainties caused by sensor noises, measurement errors, and adversarial perturbations. Although these approaches focused on improving the performance for the ego vehicle can achieve high scalability, they do not perform well in complex traffic environments that require coordination of multiple AVs~\cite{hoel2018automated}.

To address the performance limitation of the ego vehicle-based approaches, multi-agent reinforcement learning (MARL)-based solutions were proposed~\cite{ha2020leveraging,chen2021graph,zhou2022multi,hou2021decentralized,zhang2022multi}. The key focus of the MARL-based approaches is to exploit the close collaboration among multiple agents to make more efficient lane-changing decisions to improve traffic efficiency and safety.  

Hou and Graf proposed a multi-agent deep RL approach that is designed to control both the lane-changing behavior and longitudinal movement of the ego vehicle~\cite{hou2021decentralized}. The authors demonstrated that the dual control approach improved the roadway capacity, especially for highway merging and weaving areas. However, the traffic efficiency for their reward function is defined based only on the speed of the ego vehicle without considering the overall impact of a lane-changing decision on the surrounding traffic. 

Zhou \emph{et al.} created a MARL-based approach using the multi-agent advantage actor-critic network (MA2C) and a novel parameter-sharing scheme to facilitate collaboration among agents~\cite{zhou2022multi}. Noting that the reward functions of existing MARL-based approaches do not fully consider the passenger's comfort, the authors focused on designing a MARL-based approach that simultaneously improves traffic efficiency, safety, and driving comfort. However, similar to~\cite{hou2021decentralized}, their reward function does not account for the overall impact of lane-changing behavior on surrounding traffic. Furthermore, the representation of the driving comfort in their reward function is simplistic, \emph{i.e.,} determined based only on a specific threshold. 

Wang \emph{et al.} designed a MARL-based lane-changing solution without relying on vehicle-to-everything (V2X) communication~\cite{wang2021harmonious}.  The proposed approach is targeted to optimize both the individual and overall efficiency based only on the limited sensing results of individual vehicles. However, since no V2X is adopted, a vehicle can only have a narrow view, which makes it difficult, if not possible, to make a decision to optimize the overall traffic flow of a road segment while ensuring traffic safety and driving comfort.

Zhang \emph{et al.} proposed a bi-level approach where the upper level is the MARL model for making lane-change decisions, and the lower level manages potential conflicts during the implementation of the lane-change decision~\cite{zhang2022multi}. The proposed approach improves efficiency by encoding the driving intentions of surrounding vehicles into the state space. Their reward function considers both the efficiency of the ego vehicle and the impact on the overall traffic efficiency. However, this paper is based on a critical assumption that there does not exist vehicular communication and uses only the onboard sensors, therefore limiting the perception capability of the ego vehicle only to its surrounding area.

Unlike recent MARL-based approaches that overlook the impact of lane-change decisions on traffic flow in a road segment, LCS-TF aims to utilize the collaborative lane-change decisions of multiple agents to enhance the overall traffic flow while simultaneously optimizing the individual efficiency, safety, and driving comfort of the agents.




\section{Multi-Agent Deep Reinforcement Learning for Intelligent Lane-Changing}
\label{sec:proposed_approach}

\subsection{Overview}
\label{sec:overview}

Fig.~\ref{fig:overview} illustrates an operational overview of LCS-TF. LCF-TF is deployed to cover a roadway segment with mixed traffic of AVs and HVs. We assume that AVs are equipped with the V2X capability. We also assume that there exists a roadside unit (RSU) that is assigned to monitor and manage traffic on a particular roadway segment. More specifically, the RSU gathers traffic data from the roadway segment and disseminates the aggregated information to the agents within the same segment.

LCS-TF allows the RL agents to make lane-change decisions based on both the local and global traffic information, aiming to optimize the traffic flow of the roadway segment as well as the efficiency, safety, and driving comfort of the ego vehicle. More specifically, the local information is obtained from the surrounding vehicles using the ego vehicle's onboard sensors. The global information is directly received from the RSU via V2X. More details on the different kinds of local and global traffic information used by the agents to make lane-change decisions, as well as the corresponding machine learning algorithm are discussed in Section~\ref{subsec:state_space}.

\subsection{Problem Formulation}
\label{sec:problem_formulation}

The problem of making a lane-change decision for each RL agent to jointly optimize the traffic flow and the performance of the ego vehicle can be represented as a decentralized partially observable Markov decision process (decPOMDP). The decPOMDP is described by the tuple $(\{\mathcal{A}_i, \mathcal{O}_i, \mathcal{R}_i\}_{i \in \mathcal{V}}, \mathcal{T}, \mathcal{S})$ where $\mathcal{A}_i$ is an action space for an agent $i \in \mathcal{V}$, where $\mathcal{V}$ is a set of agents. $\mathcal{O}_i$ is a partial observation of the environmental state $\mathcal{S}$ for agent $i$. $\mathcal{T}$ is a set of transitional probabilities which is unknown. The goal is to find a decentralized policy $\pi_i:\mathcal{O}_i \times \mathcal{S} \rightarrow [0,1]$ for each agent $i$ to choose an action $a_t$ at time $t$ such that the expected cumulative reward is maximized. LCS-TF solves the problem based on a MARL framework for which the details are presented in subsequent sections.

\subsection{State Space}
\label{subsec:state_space}

The state space $\mathcal{O}_i$ of agent $i$ consists of both local and global traffic information. The local traffic information refers to the traffic state of the ego vehicle and its surrounding vehicles. Specifically, vehicles that fall within a range of 100m from the ego vehicle are referred to as surrounding vehicles. Therefore, the traffic state for surrounding vehicles is given as a 2D vector $N \times \mathcal{F}$, where $N$ is the number of surrounding vehicles, and $\mathcal{F}$ is the feature set. The feature set $\mathcal{F}$ of a surrounding vehicle consists of the longitudinal/lateral position, longitudinal/lateral speed, acceleration, and driver imperfection of the vehicle. In particular, driver imperfection is incorporated to take into account the driving behavior of HVs with an aim for the agents to make more efficient lane-change decisions. The traffic state for the ego vehicle includes the ego vehicle's longitudinal/lateral position, longitudinal/lateral speed, and acceleration. 

The state space $\mathcal{O}_i$ of agent $i$ also integrates global traffic information, enabling the agent to make lane-change decisions while considering the effect on traffic flow for the present road segment. The global traffic information includes the vehicle density, average vehicle speed, maximum allowed speed, and the number of lanes of the roadway segment. It also includes the mean speed and vehicle density for each lane. Additionally, the key road-specific parameters used by the agent to make lane-change decisions, including the lateral safety distance, longitudinal safety distance, and vehicle action update interval, are also included in the global traffic information. The parameter values used in this study are presented in the description of the proposed reward function.

\subsection{Action Space}
\label{subsec:action_space}

The action space $\mathcal{A}_i$ of agent $i$ consists of five different discrete lane-change and/or speed adjustment decisions as shown below.

\begin{displaymath}
	\mathcal{A}_i=\{\mbox{left, right, keep, accelerate, decelerate}\}.
\end{displaymath} 

\noindent More specifically, an agent $i$ can undertake the following actions, namely, (i) switching to the left lane, (ii) switching to the right lane, (iii) maintaining the current speed, (iv) accelerating, and (v) decelerating. In this study, the action update interval was set to 0.1s.

\subsection{Reward Function}
\label{subsec:reward_function}

The reward function $\mathcal{R}_i$ for agent $i$ is comprised of four subfunctions, which represent the traffic efficiency $r_e$, safety $r_s$, and driving comfort $r_c$, and lane change utility $r_u$ for the ego vehicle, respectively, \emph{i.e.,}

\begin{displaymath}
	R_i = w_1 r_e+w_2 r_s+w_3 r_c+w_4r_u,
\end{displaymath} 

\noindent where $w_1$, $w_2$, $w_3$ and $w_4$ are weighting parameters.

\textbf{Efficiency Reward:} The efficiency reward function $r_e$ not only incentivizes RL agents to improve their own speed but also promotes the flow of traffic within the road segment. More specifically, the efficiency reward function $r_e$ is defined as follows.

\begin{displaymath}
	r_e  = g_e + l_e,
\end{displaymath}

\noindent where $g_e$ denotes the global efficiency, and $l_e$ denotes the local efficiency. Global efficiency represents the effect of the ego vehicle's action on the traffic flow for the road segment. More specifically, the global efficiency $g_e$ is defined as follows. 

\begin{displaymath}
	g_e = 
	\begin{dcases}
		- \frac{v_e - v_{max}}{v_{max}}, & \text{if } v_e > v_{max}\\
		\frac{v_e - v_{min}}{v_{min}}, & \text{if } v_{min} \le v_e \le v_{max}\\
		- \frac{v_{min} - v_e}{v_{min}},              & \text{otherwise,}
	\end{dcases}
\end{displaymath}

\noindent where $v_e$ is the average speed of vehicles in the road segment. $v_{min}$ and $v_{max}$ are the minimum average speed and maximum speed limit defined for the road segment, respectively. A positive reward is granted when the ego vehicle's lane-change decision leads to an improvement in $v_{e}$, provided that it lies between $v_{min}$ and $v_{max}$. Otherwise, a negative reward is issued. When computing the reward, we ensure that it is proportional to the difference between $v_e$ and $v_{min}$ (or $v_{max}$).

The local efficiency $l_e$ is determined based on the speed of the ego vehicle in comparison with $v_{min}$ and $v_{max}$, \emph{i.e.,}

\begin{displaymath}
	l_e = 
	\begin{dcases}
		- \frac{v_{ego} - v_{max}}{v_{max}}, & \text{if } v_{ego} > v_{max}\\
		\frac{v_{ego} - v_{min}}{v_{min}}, & \text{if } v_{min} \le v_{ego} \le v_{max}\\
		- \frac{v_{min} - v_{ego}}{v_{min}},              & \text{otherwise,}
	\end{dcases}
\end{displaymath}

\noindent where $v_{ego}$ is the speed of the ego vehicle. The way how the local efficiency reward is computed is similar to that of the global efficiency reward.


\begin{algorithm}[t]
	\begin{algorithmic}[1]
		\REQUIRE {Episodes \textbf{\textit{E}}, Episode Duration \textbf{\textit{T}}, discount factor $\gamma$, Memory Size $\textit{M}$, Learning start size $n$, target network update frequency $t^\prime$}
		\ENSURE{Action, $a \in$ \{right, left, accelerate, decelerate, keep\}}
		\STATE{Initialize reply memory $\textit{B}$ to size $\textit{M}$}
		\STATE {Initialize action-value function $Q$ with random weights}
		\STATE {Initialize target network parameters with online network weights}
		\FOR {$episode = 1 : \textit{E}$}
		\STATE Get initial observation $s_1$
		\FOR {$t = 1 : \textit{T}$}
		\FOR{$i\in RL$}
		\STATE{select a random action $a_t^i$ with probability $\epsilon$}
		\STATE{otherwise action $a_t^i = \text{max}_aQ^*(s_t^i,a;\theta_i)$}
		\STATE{apply action $a_t^i$, observe reward $r_t^i$ and next observation $s_{t+1}^i$}
		\STATE{store transition $(s_t^i, a_t^i, r_t^i, s_{t+1}^i)$ in $\textit{B}$}
        \ENDFOR
		\STATE \hskip 0.0em {sample random minibatch of transitions $(s_k, a_k, r_k, s_{k+1})$ from $\textit{B}$}
		\STATE \hskip 0.0em {compute 
			\begin{equation*}
				y_k = \begin{cases}
					r_k & \text{if episode ends}\\
					r_k+\gamma \text{max}_{a^\prime}(s_{k+1},a^\prime;\theta) & \text{otherwise}
				\end{cases}
			\end{equation*}
		}
		\STATE \hskip 0.0em{compute loss $(y_k-Q(s_j,a;\theta))^2$}
		\STATE{\hskip 0.0em{adjust online network weights using eq (\ref{MADQNLC})}}
		\STATE \hskip 0.0em{update target network weights at every $t^\prime$}
		\ENDFOR
		\ENDFOR
	\end{algorithmic}
	\caption{Multi-Agent DQN (MADQN) Lane Change Decision}
	\label{MADQNLC}
\end{algorithm}

\textbf{Safety Reward:} The safety reward function $r_s$ is designed to promote the safety of both the ego vehicle and the surrounding vehicles. More specifically, the lateral and longitudinal safety of the ego vehicle and the potential collision with the surrounding vehicles are incorporated in the safety reward function, \emph{i.e.,}

\begin{displaymath}
	r_s = l_{lon}+l_{lat}+l_{col},
\end{displaymath}

\noindent where the longitudinal safety component $l_{lon}$ of the reward function ensures that the agent maintains the minimum safety distance while executing lane changes, \emph{i.e.,} 
\begin{displaymath}
	l_{lon} = 
	\begin{dcases}
		\frac{d_{long} - T_{minGap}}{T_{minGap}}, & \text{if $d_{long}$} \le \text{$T_{minGap}$(2.5m)}\\
		0,              & \text{otherwise}
	\end{dcases}
\end{displaymath}
where $d_{long}$ is the longitudinal distance, and $T_{minGap}$ is the longitudinal safety threshold.

\noindent The lateral safety component $l_{lat}$ ensures that a safe lateral gap is maintained while executing lane changes, \emph{i.e.,}
\begin{displaymath}
	l_{lat} = 
	\begin{dcases}
		\frac{d_{lat}-T_{lat}}{T_{lat}}, & \text{if $d_{lat}$} \le \text{ $T_{lat}$ (10m)}\\
		0,              & \text{otherwise}
	\end{dcases}
\end{displaymath}
where $d_{lat}$ is the lateral distance, and $T_{lat}$ is the lateral safety threshold.

\noindent The collision component $l_{col}$ imposes a significant penalty on the agent in the event of a collision with surrounding vehicles, \emph{i.e.,}

\begin{displaymath}
	l_{col} = 
	\begin{dcases}
		-5, & \text{if collide with other vehicle}\\
		0,              & \text{otherwise}
	\end{dcases}
\end{displaymath}

\textbf{Driving Comfort Reward:} We employ an existing methodology to quantify driving comfort~\cite{zhu2020safe,jacobson1980models,das2021saint}. Specifically, the driving comfort reward function is designed based on the change rate of acceleration as follows.

\begin{displaymath}
	r_{c} = -\frac{\Delta a}{jerk_{max}^2},
\end{displaymath}

\noindent where $jerk$ is calculated based on the acceleration range. For example, in our simulation, an agent makes a decision every 0.1s, and the acceleration range is between 2.6 $m/s^2$ to -2.6 $m/s^2$. Hence, the greatest possible jerk $jerk_{max}$ is (2.6-(-2.6))/0.1=52.  

\textbf{Lane Change Utility Reward:} Although the reward function $R_i$ is designed to maximize traffic efficiency, safety, and comfort, we observe that agents at times make invalid lane-change decisions, especially during the initial period of training. To facilitate the training process and achieve better performance, we incorporate the lane change utility reward function $r_u$. It ensures that an agent is penalized with a negative reward when it makes an invalid lane-change decision which includes (1) a lane-change decision to switch to the left lane when the vehicle is already in the leftmost lane (-0.5), (2) a lane-change decision to switch to the right lane when the vehicle is already in the rightmost lane (-0.5), (3) a lane-change decision when there is no traffic in front of the vehicle (-5), (4) a lane-change decision when the leading vehicle is moving faster than the agent (-0.5), and (5) a lane-change decision when the leading vehicle in the target lane is moving slower than the agent (-0.5). It's worth noting that the reward values are established via an extensive trial-and-error process.



\subsection{Multi-Agent DQN Algorithm}

We propose a multi-agent DQN algorithm designed to identify the optimal policy that facilitates the efficient coordination of multiple agents in maximizing the cumulative reward, based on their partial observations of the environment. We adopt the centralized training and decentralized execution paradigm where each RL agent shares the same network architecture and parameters~\cite{lowe2017multi,tan1993multi}. In this shared learning architecture, each RL agent senses surrounding traffic conditions as observations and performs optimal longitudinal or lateral actions based on the current observations. Furthermore, each agent stores its experiences in shared memory, which contributes to updating policy network weights by minimizing the loss function at each gradient step based on Eq.~\ref{eq:gradient_update}; as a result, other agents benefit from performing optimal decisions using updated policy networks. 
\begin{displaymath}
    y_i = r+\gamma\text{max}_{a^\prime}Q(s^\prime,a^\prime;{\theta_i}^\prime)
\end{displaymath}
\begin{displaymath}
	\label{eq:gradient_update}
	\nabla_{\theta_i} L_i(\theta_i) = \mathbb{E}_{s,a\sim\rho(.);s^\prime}\Bigr[\Bigr(y_i-Q(s,a;\theta_i)\Bigr)\nabla_{\theta_i}Q(s,a;\theta_i)\Bigr]
\end{displaymath}


\noindent where $Q(s,a;\theta_i) (s \in \mathcal{S}, a \in \mathcal{A}$) denotes the Q-network with weights $\theta_i$ for agent $i$, which is used to estimate the action-value function. ${\theta_i}^\prime$ is the parameters in the previous iteration for agent $i$. $\rho(s,a)$ is a probability distribution over observations $s$ and actions $a$. Moreover, each agent stores its experiences in the shared memory so that other agents can benefit. The algorithm \ref{MADQNLC} outlines the pseudocode for our multi-agent lane change decision-making strategy. 

At every simulation step, all agents perform actions simultaneously following the $\epsilon-$greedy policy~\cite{wunder2010classes} and store their transitions in the shared replay buffer. We trained the policy network by taking a mini-batch of uniform random samples of transitions from the replay buffer at every simulation second to optimize the network parameters so that agents can make better decisions.


\begin{table}[ht!]
	\centering
	\begin{tabular}{|ll|}
		\hline
		\multicolumn{2}{|c|}{Neural Network Hyperparameters}              \\ \hline
		\multicolumn{1}{|l|}{minibatch size}             & 32          \\ \hline
		\multicolumn{1}{|l|}{reply buffer size}             & 400k        \\ \hline
		\multicolumn{1}{|l|}{discount factory}              & 0.99        \\ \hline
		\multicolumn{1}{|l|}{exploration start}             & 1.0         \\ \hline
		\multicolumn{1}{|l|}{exploration end}               & 0.1         \\ \hline
		\multicolumn{1}{|l|}{exploration decay}             & 0.99985     \\ \hline
		\multicolumn{1}{|l|}{learning rate}                 & 0.00025      \\ \hline
		\multicolumn{1}{|l|}{target update frequency}       & 10 episodes \\ \hline
		\multicolumn{1}{|l|}{State space size}              & 40          \\ \hline
		\multicolumn{1}{|l|}{action space size (discrete)} & 5           \\ \hline
		\multicolumn{1}{|l|}{activation function (hidden/output)} & ReLU/linear                                                                            \\ \hline
		\multicolumn{1}{|l|}{Loss function } & Huber Loss                                                                            \\ \hline
		\multicolumn{1}{|l|}{network size}                        & \begin{tabular}[c]{@{}l@{}}4 hidden layer with \\ 32, 64, 64, 512 hidden units\end{tabular} \\ \hline
	\end{tabular}
	\caption{Optimal hyperparameter values for LCS-TF.}
	\label{tab:lc_hyperparameters}
\end{table}

\section{Simulation Results}
\label{sec:experiments}

In this section, we present the simulation results. Section~\ref{sec:experimental_settings} introduces the simulation settings. Section~\ref{sec:training} discusses the details of training the model. We then analyze the traffic efficiency, traffic safety, and driver comfort of the proposed solution in Sections~\ref{sec:traffic_efficiency},~\ref{sec:traffic_safety}, and~\ref{sec:driver_comfort}, respectively. 

\subsection{Simulation Settings}
\label{sec:experimental_settings}

We used  a traffic simulator called SUMO~\cite{lopez2018microscopic} to implement LCS-TF and two main types of approaches, namely NP-MARL, \emph{i.e.,} a MARL-based approach with no V2X support~\cite{zhang2022multi}, and LSC-MARL which uses V2X but has limited functionalities for improving traffic safety and driving comfort~\cite{zhou2022multi}. The MARL models were created based on Keras and Tensorflow~\cite{abadi2016tensorflow} and were interfaced with SUMO via Traffic Control Interface (TraCI)~\cite{wegener2008traci}. A workstation equipped with the Intel Xeon Gold 5222 Processor, NVIDIA® RTX™ A4000, and 48GB RAM running on Windows 11 OS was used to train and test the MARL models. 

We consider a 3.5km-long, 5-lane road segment. The initial 500m of the road segment was allocated for vehicle injection and acceleration for both AVs and HVs. The following 3km of the road segment was where RL agents perform lane-changes. The speed limit for the road segment was set to 75mph. Vehicles were randomly injected at a rate of 2,160 veh/h for which the proportion of AVs was varied for evaluation purposes. The initial speed of the vehicles was 45mph, and the range of acceleration for each vehicle was set to -2.6$m/s^2$ to 2.6$m/s^2$. The intelligent driver model (IDM) ~\cite{treiber2000congested, lopez2018microscopic} was used to simulate the driving behavior of HVs, and a widely-used lane-change model~\cite{semrau2016simulation} was employed to simulate the lane-changing behavior of HVs. For a more realistic simulation, four different types of HVs with different body lengths and driving imperfections were used.

The average speed of all vehicles in the road segment was used as the main metric to represent traffic flow. The rate of collisions that occur while executing lane changes was utilized as a key metric for evaluating driving safety. Moreover, we assessed driver comfort by quantifying changes in the vehicle's acceleration and deceleration. 

\subsection{Training MARL Model}
\label{sec:training}

\begin{figure}[h]
	\centering
	\includegraphics[width=.99\columnwidth]{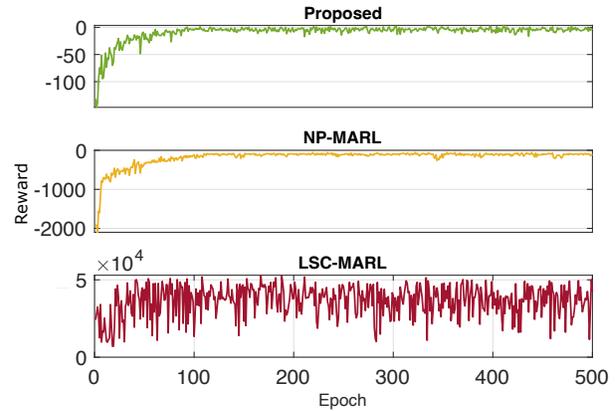}
	\caption {Convergence of the reward function of three different MARL models.}
	\label{fig:reward}
\end{figure}

In this section, we present the details regarding training the MARL model. The optimal values for the hyperparameters used for training the model were identified based on the grid search method ~\cite{liashchynskyi2019grid} (See Table 1). The MARL model was trained until the network converges. As depicted in Fig.~\ref{fig:reward}, our MARL model (as well as that for NP-MARL and LSC-MARL)  converged within 500 epochs. In our simulation setting, the simulation update interval was 0.1s, and the duration of an epoch was 5 minutes of simulation steps. Agents updated the model parameters every 10 simulation steps which are equivalent to 1s.

\subsection{Traffic Efficiency}
\label{sec:traffic_efficiency}

\begin{figure}[h]
	\centering
	\includegraphics[width=.99\columnwidth]{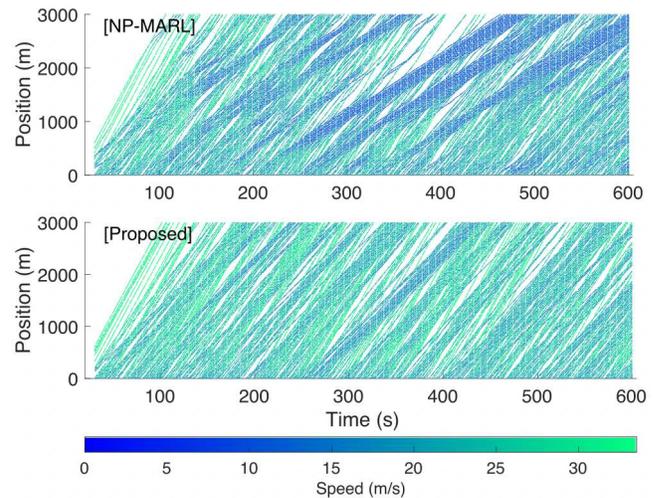}
	\caption {A space-time diagram representing traffic efficiency for NP-MARL and LCS-TF.}
	\label{fig:global_efficiency}
\end{figure}

This section presents simulation results related to traffic efficiency. A space-time diagram depicted in Fig.~\ref{fig:global_efficiency} provides a general view of how NP-MARL and LCS-TF perform in terms of traffic efficiency. As illustrated, NP-MARL led to significant speed reductions throughout the highway segment. This can be attributed to the fact that NP-MARL failed to consider the broader impact of lane-change decisions on global traffic efficiency. Conversely, the figure provides clear evidence that the LCS-TF's coordinated lane-change decisions for agents resulted in a significant reduction of overall vehicle speed throughout the entire segment.

\begin{figure}[h]
	\centering
	\includegraphics[width=.99\columnwidth]{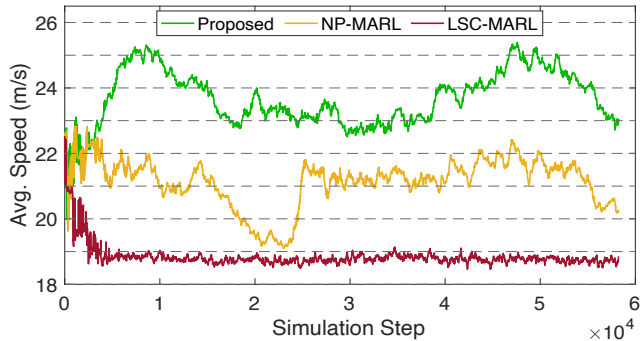}
	\caption {The average speed changes over time for LCS-TF, NP-MARL, and LSC-MARL.}
	\label{fig:traffic_flow}
`\end{figure}

We then examined how the average speed changes over time for the three MARL models. Fig.~\ref{fig:traffic_flow} depicts the results. It was observed that despite the random driving behavior of HVs, overall, LCS-TF achieved a significant improvement in the average speed compared with the two other solutions. More specifically, LCS-TF improved the average speed by up to 12\% and 25.7\% compared with NP-MARL and LSC-MARL, respectively. Interestingly, compared to the other solutions, we observed agents refraining from executing lane changes even though immediate speed improvement was expected. Instead, those agents allowed surrounding vehicles to increase their speed, leading to the overall improvement in traffic flow for the road segment.


\begin{figure}[h]
	\centering
	\includegraphics[width=.99\columnwidth]{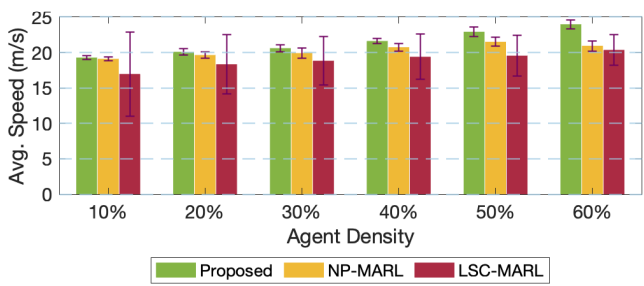}
	\caption {Traffic efficiency with varying agent densities.}
	\label{fig:traffic_flow_agent_density}
\end{figure}

We also evaluated traffic efficiency by varying the agent density. Here the agent density refers to the percentage of RL agents out of all vehicles injected into the road segment. Fig.~\ref{fig:traffic_flow_agent_density} shows the results. As depicted, traffic flow improved for all models, as the agent density increased. We made two key observations. First, LCS-TF consistently outperformed the other two models in terms of traffic efficiency, regardless of the agent density. More specifically, on average, LCS-TF achieved 5.3\% and 13.2\% higher average speed compared with that for NP-MARL and LSC-MARL, respectively. Notably, LCS-TF exhibited a significantly greater performance improvement at higher agent densities, as compared to the other two models. Specifically, for an agent density of 60\%, LCS-TF achieved a higher performance gain by 14.55\% and 17.78\% compared to NP-MARL and LSC-MARL. In contrast, for an agent density of 10\%, the performance gain was 0.9\% and 13.7\% higher than that of NP-MARL and LSC-MARL, respectively.

\subsection{Driving Safety}
\label{sec:traffic_safety}

\begin{figure}[h]
	\centering
	\includegraphics[width=.99\columnwidth]{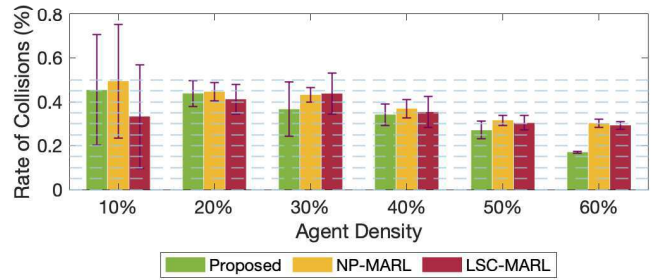}
	\caption {Driving safety with varying agent densities.}
	\label{fig:safety_agent_density}
\end{figure}

This section evaluates the performance of LCS-TF focusing on driving safety in comparison with NP-MARL and LSC-MARL. In particular, we measured the rate of collisions to represent driving safety. Fig.~\ref{fig:safety_agent_density} depicts the results. As shown, in general, driving safety for all three models improved as the agent density increased. A noteworthy observation was that driving safety for LCS-TF more sharply improved compared to the other two models. More specifically, for LCS-TF, the rate of collisions decreased by 62.5\% as the agent density increased from 10\% to 60\%; in contrast, NP-MARL and LSC-MARL achieved improvements of only 38.9\% and 12.2\%, respectively. We also noted that under low agent density, all MARL models exhibited unstable performance. On the other hand, more stable performance was achieved as the agent density increased.

\subsection{Driver Comfort}
\label{sec:driver_comfort}

This section presents simulation results on driving comfort. The driving comfort was measured based on the $jerk$ as defined in the driving comfort reward. Note that the value of $jerk$ is always negative, and a higher jerk value means better driving comfort.

\begin{figure}[h]
	\centering
	\includegraphics[width=.99\columnwidth]{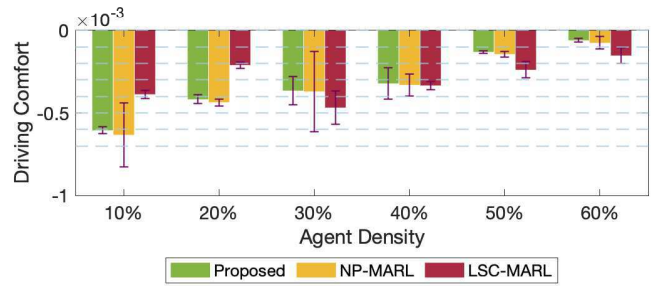}
	\caption {Driving comfort with varying agent densities.}
	\label{fig:driving_comfort_agent_density}
\end{figure}

Fig.~\ref{fig:driving_comfort_agent_density} depicts the results. In general, as the agent density increased, there were more vehicles executing lane changes with greater care. Therefore, with the higher agent density, we observed the increasing trend of driving comfort for all three models. While improvement in driving comfort was achieved for all models, a notable observation was that LCS-TF achieved the highest driving comfort compared to the other two solutions. Specifically, LCS-TF outperformed by up to 19.4\% and 60.2\% in terms of driving comfort in comparison with that for NP-MARL and LSC-MARL, respectively.

\section{Conclusion}
\label{sec:conclusion}

We have presented LCS-TF: a multi-agent deep reinforcement learning-based intelligent lane-change system for autonomous vehicles. The design of LCS-TF aims to achieve joint optimization of both the ego vehicle's performance and the overall traffic flow for a given road segment. To achieve the goal, a novel state space design was introduced that incorporates the local traffic information obtained from surrounding vehicles along with the global traffic information collected from the RSU assigned to manage the road segment. We also presented a novel reward function to ensure effective lane-change decisions that improve traffic flow and optimize the ego vehicle's performance. Additionally, a multi-agent deep Q-network (DQN) algorithm was developed to identify the optimal policy for each agent to effectively collaborate in performing lane-change maneuvers. LCS-TF's effectiveness was assessed via extensive simulations and compared against the two main types of existing MARL models. The results indicated that LCS-TF outperforms existing MARL models in all aspects of traffic efficiency, driving safety, and driver comfort. Our future work is to extend the scalability of LCS-TF. Although the current version concentrates on optimizing the performance for a specific road segment, we aim to accomplish multiple road segment-level coordination, examine a more large-scale impact of LCS-TF, and develop the next-generation intelligent lane-change system for AVs.

\bibliographystyle{IEEEtran}
\bibliography{mybibfile}

\end{document}